\documentclass{article}

\PassOptionsToPackage{numbers, compress}{natbib}
\usepackage[final]{nips_2018}

\usepackage[utf8]{inputenc} 
\usepackage[T1]{fontenc}    
\usepackage{url}            
\usepackage{booktabs}       
\usepackage{amsfonts}       
\usepackage{nicefrac}       
\usepackage{microtype}      
\usepackage{color}
\usepackage{amsmath}        
\usepackage[colorinlistoftodos]{todonotes} 
\usepackage{soul}           
\usepackage[
bookmarksopen,
breaklinks=true,
colorlinks=true,
citecolor=black,
urlcolor=blue]{hyperref}
\usepackage[export]{adjustbox}
\usepackage{subfigure}
\usepackage{graphicx}
\usepackage{booktabs}
\usepackage{breakurl}
\usepackage{siunitx}

\usepackage[compact]{titlesec}
\usepackage{adjustbox}

\title{A large-scale crowd-sourced analysis of abuse against women journalists and politicians on Twitter}

\author{
    Laure Delisle\thanks{Equal contribution.~\textsuperscript{$\dagger$}{\texttt{\{laure,freddie,km,archy,julien\}@elementai.com};~\textsuperscript{$\ddagger$}\texttt{milena.marin@amnesty.org}}} ~\footnotemark[2]{} \\
    Element AI
    \And
    Alfredo Kalaitzis\footnotemark[1]{} ~\footnotemark[2]{} \\
    Element AI
    \And
    Krzysztof Majewski\footnotemark[2]{} \\
    Element AI
    \AND
    Archy de Berker\footnotemark[2]{} \\
    Element AI
    \And
    Milena Marin\footnotemark[3]{} \\
    Amnesty International
    \And
    Julien Cornebise\footnotemark[2]{} \\
    Element AI 
}

\begin{document}
\maketitle
\vspace{-2em}
\begin{abstract}
    \vspace{-1em}
    We report the first, to the best of our knowledge, hand-in-hand collaboration between human rights activists and machine learners, leveraging crowd-sourcing to study online abuse against women on Twitter.
    On a technical front, we carefully curate an unbiased yet low-variance dataset of labeled tweets, analyze it to account for the variability of abuse perception, and establish baselines, preparing it for release to community research efforts.
    On a social impact front, this study provides the technical backbone for a media campaign aimed at raising public and deciders' awareness and elevating the standards expected from social media companies. 
\end{abstract}

\section{Introduction}  \label{sec:intro}
Social media platforms have become a critical space for women and marginalized groups to express themselves at an unprecedented scale.
Yet a stream of research by Amnesty International~\cite{Dhrodia:realtoll17, AmnestyToxicTwitter2018} showed that many women are subject to targeted online violence and abuse, which denies them the right to use social media platforms equally, freely, and without fear.
Being confronted with toxicity at a massive scale leaves a long-lasting effect on mental health, sometimes even resulting in withdrawal from public life altogether \cite{ukgov:intimidation2017}.
A first smaller-scale analysis of online abuse against women UK Members of Parliament (MPs) on Twitter~\cite{Dhrodia:womenmps17, Katia2017} proved the impact such targeted campaigns can have: it contributed to British Prime Minister Theresa May publicly calling out the impact of online abuse on democracy~\cite{MayAbuseCallout}.

This laid the groundwork for the larger-scale \textit{Troll Patrol} project that we present here: a joint effort by human rights researchers and technical experts to analyze millions of tweets through the help of online volunteers.
Our main research result is the development of a dataset that could help in developing tools to aid online moderators.
To that end, we
\emph{i)} Designed a large, enriched, yet unbiased dataset of hundreds of thousands of tweets;
\emph{ii)} Crowd-sourced its labeling to online volunteers;
\emph{iii)} Analyzed its quality via a thorough agreement analysis, to account for the personal variability of abuse perception; \emph{iv)} Compared multiple baselines with the aim of classifying a larger dataset of millions of tweets.
Beyond this collaboration, this should allow researchers worldwide to push the envelope on this very challenging task -- one of many in natural language understanding~\cite{gonzalez:sarcasm2011}.

The social impact Amnesty International is aiming for is ultimately to influence social media companies like Twitter into increasing investment and resources -- under any form -- dedicated to tackling online abuse against women.
With this study, we contribute to this social impact by providing the research backbone for a planned media campaign in November 2018.

\section{Crowd-sourcing an importance-sampled enriched set}  \label{sec:data_collection}
Core to this study is the careful crafting of a large set of tweets followed by a massive crowd-sourced data labeling effort.

\textbf{Studied population}:
We selected $778$ women politicians and journalists with an active, non-protected Twitter account, with fewer than 1 million followers, including most women British MPs and all US Congresswomen, and journalists from a range of news organizations representing a diversity of media bias.
Full details are in Appendix~\ref{sec:population}. 

\textbf{Tweet collection}: $14.5$M tweets mentioned at least one woman of interest during 2017.
We obtained a subset of $10$K per day sampled uniformly from Twitter's Firehose, minus tweets deleted since publication, totaling $2.2$M tweets.

\textbf{Pre-labeling selection}:
Taking into account the average labeling time per tweet from a pilot study, the expected duration of the campaign, and the expected graders' engagement, we targeted labeling at most $275$K tweets in triplicate.
We first selected $215$K tweets, correcting the $10$K daily cap using per-day stratified sampling proportional to each day's actual volume.
While this sample is statistically representative of the actual tweet distributions, its class imbalance would induce high variance into any estimator, and waste the graders' engagement.
We therefore enriched the dataset with $60$K tweets pre-filtered through the Naive-Bayes classifier pre-trained in~\cite{Katia2017}.
To maintain statistical non-bias, we keep track of the importance sampling weights.
    
\begin{figure}[t!]
	\centering
    \hspace{-.03\linewidth}
    \subfigure{
        \adjincludegraphics[height=2.0cm, trim={0 0 {.23\width} 0}, clip]
            {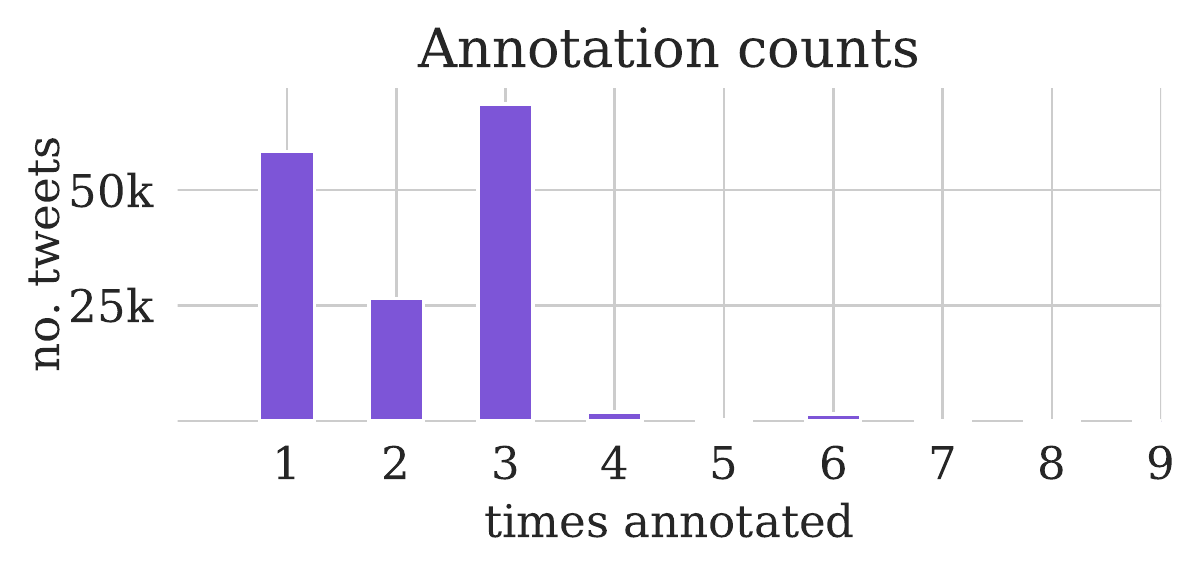}
    	\label{fig:bar_annot_counts}
    }
    \hspace{-.02\linewidth}
    \subfigure{
    	\includegraphics[height=2.0cm]{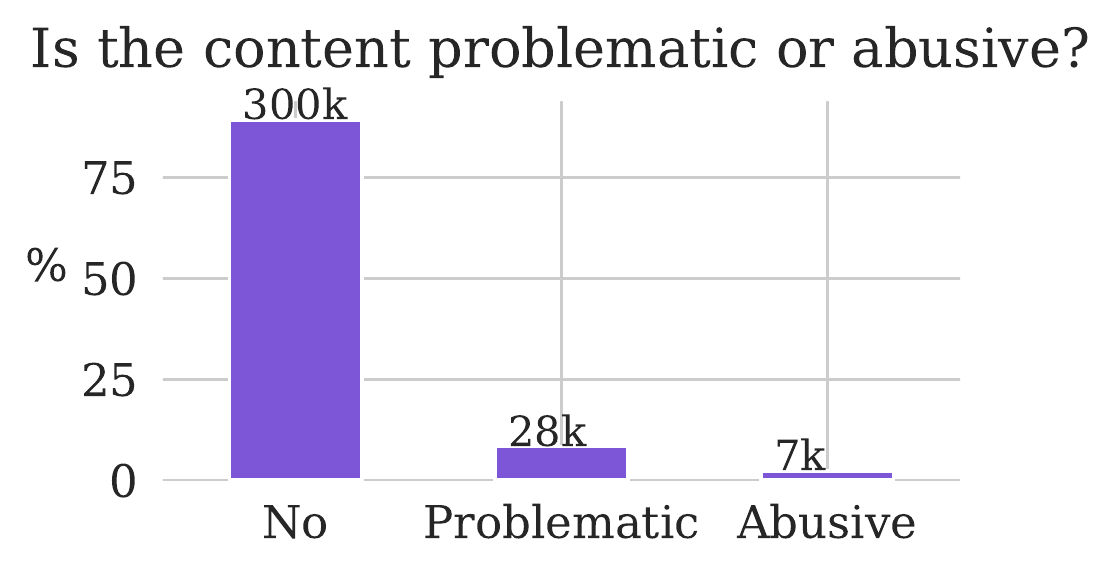}
    	\label{fig:bar_contain_abuse}
    }
    \hspace{-.03\linewidth}
    \subfigure{
        \includegraphics[height=2.0cm]{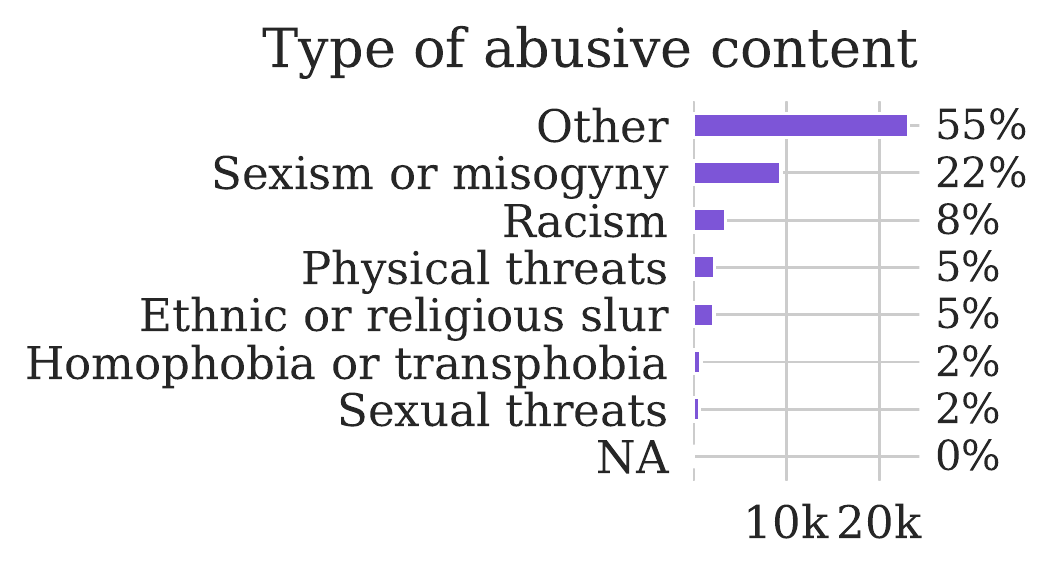}
        \label{fig:bar_type_abuse}
	}
	\hspace{-.03\linewidth}
    \subfigure{
        \includegraphics[height=2.0cm]{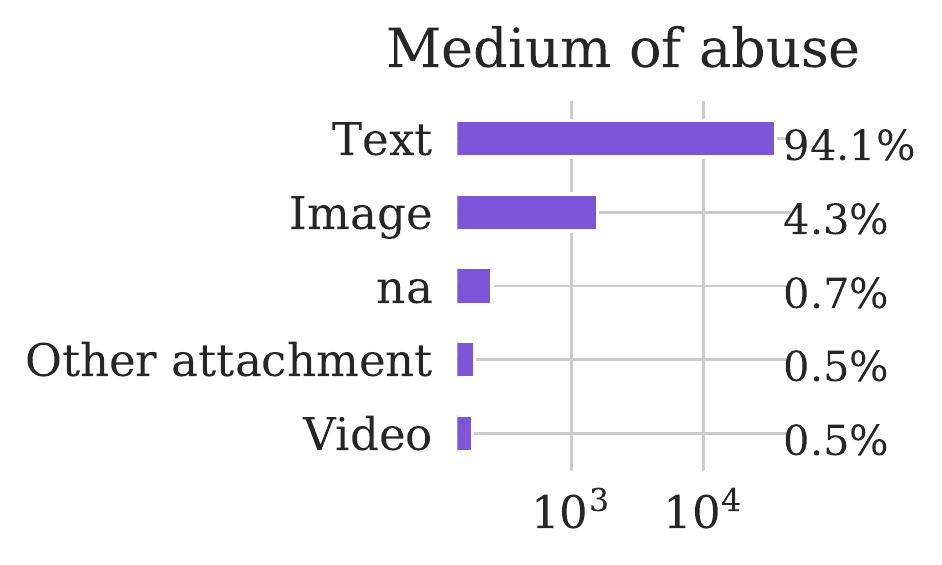}
        \label{fig:bar_medium_abuse}
	}
    \vspace{.1cm}
	\caption{ \label{fig:bars}
	    \small
	    Left to right:
    	Distribution of annotations-per-tweet: To analyze agreement, we used only tweets annotated more than twice ($\sim$73k);
    	Values of \texttt{Contain Abuse} are ordinal;
    	\texttt{Type} and \texttt{Medium} conditioned on \texttt{Contain Abuse} $\neq$ \texttt{No}:
    	the majority of abuse is not easily classified, and the vast majority of abuse is textual.
    }
\end{figure}

\textbf{Volunteers labeling via crowd-sourcing}:
Finally, these tweets, properly randomized, were deployed through Amnesty Decoders, the micro-tasking platform  based on Hive~\cite{HiveCrowdsourcing} and Discourse~\cite{Discourse} where Amnesty International engages digital volunteers  (mostly existing members and supporters) in human rights research.
Great effort was put into designing a user-friendly, interactive interface, accessible at~\cite{TrollPatrol} -- see Appendix~\ref{sec:screenshots} for screenshots.
After a video tutorial, volunteers were shown an anonymized tweet from the randomized sample, then were asked multiple-choice questions:
1) ``Does the tweet contain problematic or abusive content?'' (\textit{No}, \textit{Problematic}, \textit{Abusive}).
Unless their answer was \textit{No}, the follow-up questions were ``What type of problematic or abusive content does it contain?'' (at least one of six) and (optional question) ``What is the medium of abuse?'' (one of four). See Fig. \ref{fig:bars} for details and summary statistics.  At all times they had access to definitions and examples of abusive and problematic content, and the typologies thereof -- see Appendix \ref{sec:def_abuse}.

By August 2018, $157$K unique tweets containing $167$K mentions of the studied individuals had been categorized at least once, totalling $337$K labels, thanks to the contribution of $4,537$ online volunteers.

\textbf{Experts labeling}: In addition to engaging digital volunteers, Amnesty also asked three experts (Amnesty's researcher on online abuse against women, Amnesty's manager of the \textit{Troll Patrol} project and an external expert in online abuse) to label a sub-set of $568$ tweets. Those tweets were sampled from tweets labeled by exactly three volunteers as of June 8, 2018. To ensure low variance in the estimates, we once again used importance sampling, inflating the proportion of potentially abusive tweets by sampling 500 tweets uniformly from those labeled as ``Abusive'' by the Naive-Bayes classifier mentioned in~\cite{Katia2017} and ``Basic Negative'' by Crimson Hexagon's sentiment analysis (our Firehose access provider), and 500 tweets uniformly sampled on the remainder.
    
\textbf{Re-weighting after importance sampling}:
To ensure that any inference or training based on the enriched sample is representative of the Twitter distribution, we use importance sampling to re-weight the tweets in the empirical distribution.
The weights are defined as the ratio of the target distribution (as estimated by the daily counts) and the enriched distribution
-- see Appendix \ref{sec:reweight} for the full derivation of the weights.

\section{Analysis and generalization}  \label{sec:analysis}

\subsection{Agreement analysis}
We quantified the agreement among raters -- within crowd and within experts -- using \textit{Fleiss' kappa} ($\kappa$),
a statistical measure of inter-rater agreement \cite{Fleiss:kappa71}.
$\kappa$ is designed for \textit{nominal} (non-ordinal categorical) variables, e.g.\ Fig \ref{fig:bar_type_abuse}, whereas in ordinal variables $\kappa$ tends to underestimate the agreement because it treats the disagreement between \textit{Problematic} <> \textit{Abusive} the same as \textit{No} <> \textit{Abusive}.
We also use the \textit{intra-class correlation} (ICC) \cite{Shrout:icc79} for ordinal categorical annotations, like \texttt{Contains Abuse}: \textit{No} < \textit{Problematic} < \textit{Abusive}.
We define $\kappa$ and $\textrm{icc}$ below, and further explain in Appendix~\ref{appendix:agreement}.

\vspace{-.5cm}
\begin{figure}[ht]
	\centering
	\hspace{-.04\linewidth}
    \subfigure{
    	\includegraphics[width=.24\linewidth]{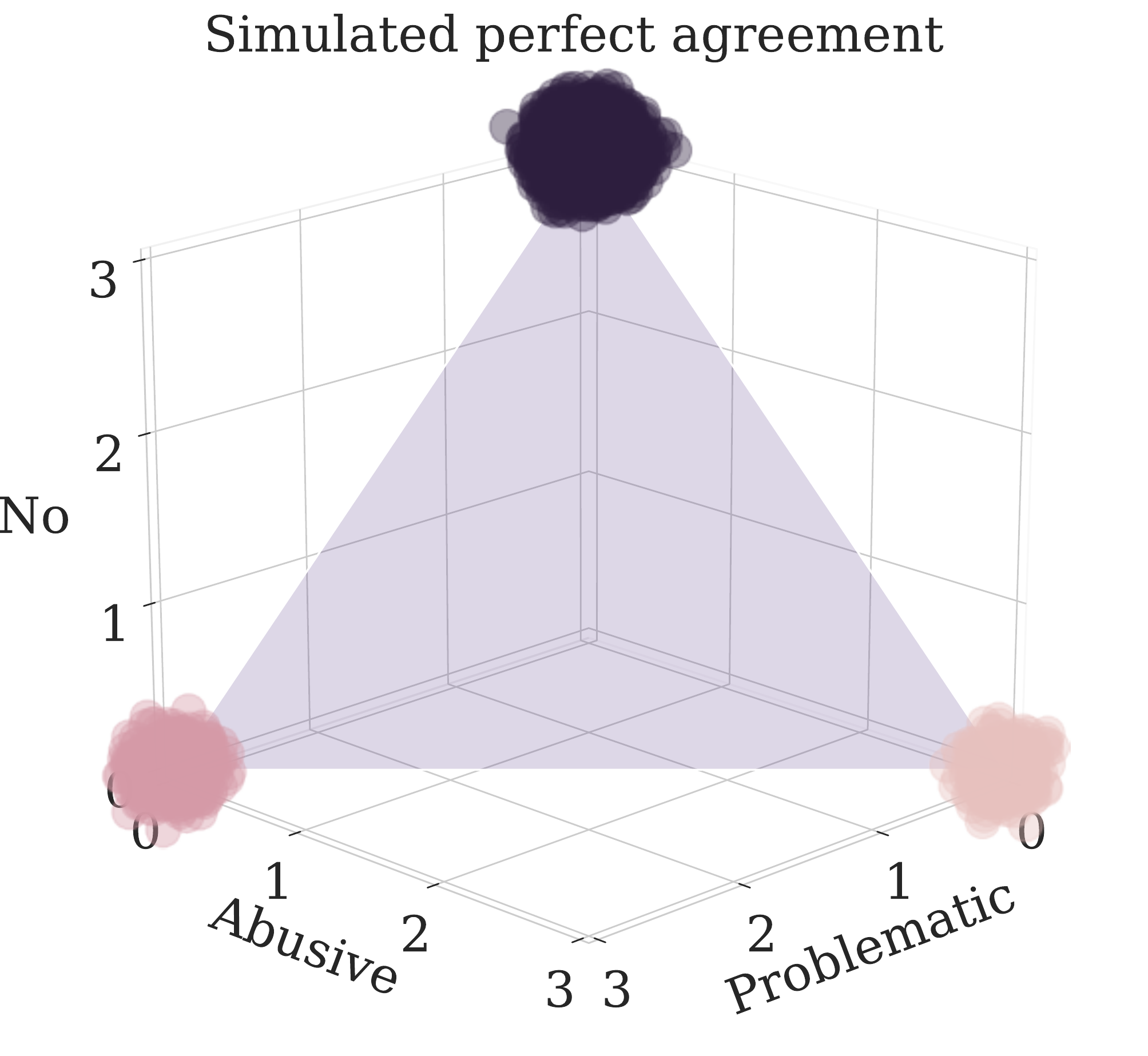}
    	\label{fig:simplex:perfect}
    }
    \hspace{-.04\linewidth}
    \subfigure{
        \includegraphics[width=.24\linewidth]{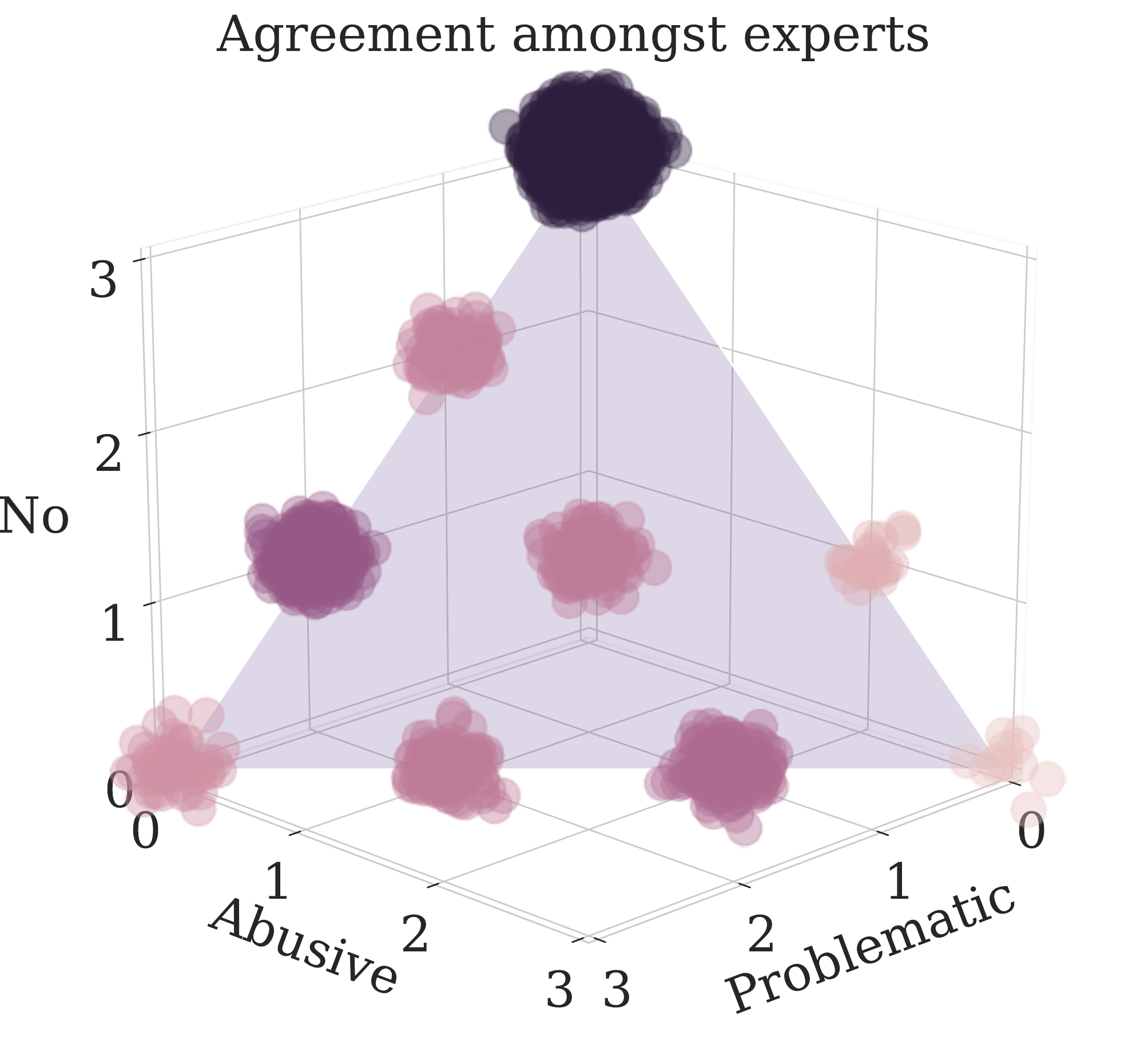}
        \label{fig:simplex:experts}
	}
	\hspace{-.04\linewidth}
	\subfigure{
        \includegraphics[width=.24\linewidth]{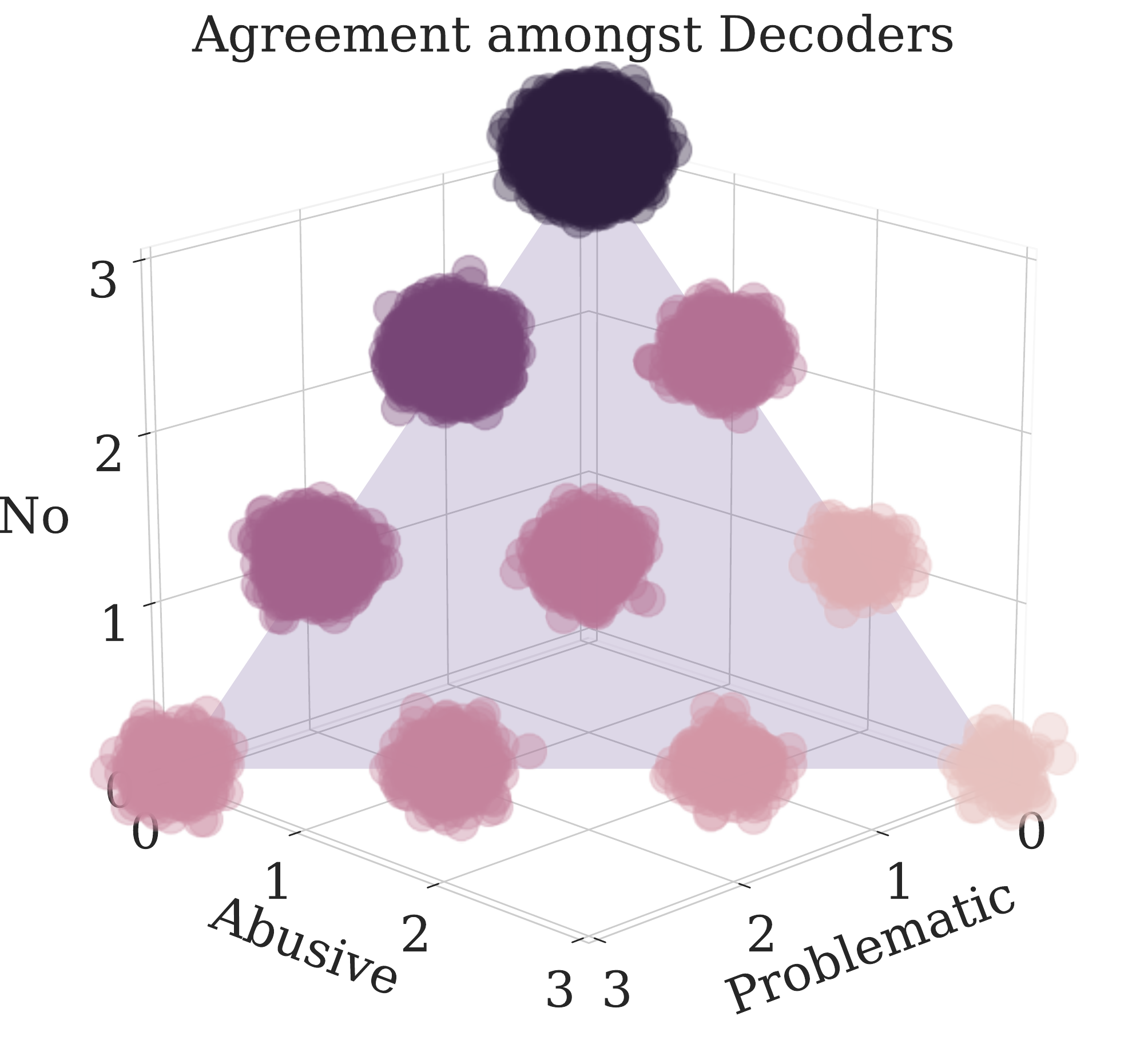}
        \label{fig:simplex:decoders}
	}
	\hspace{-.04\linewidth}
	\subfigure{
        \includegraphics[width=.26\linewidth]{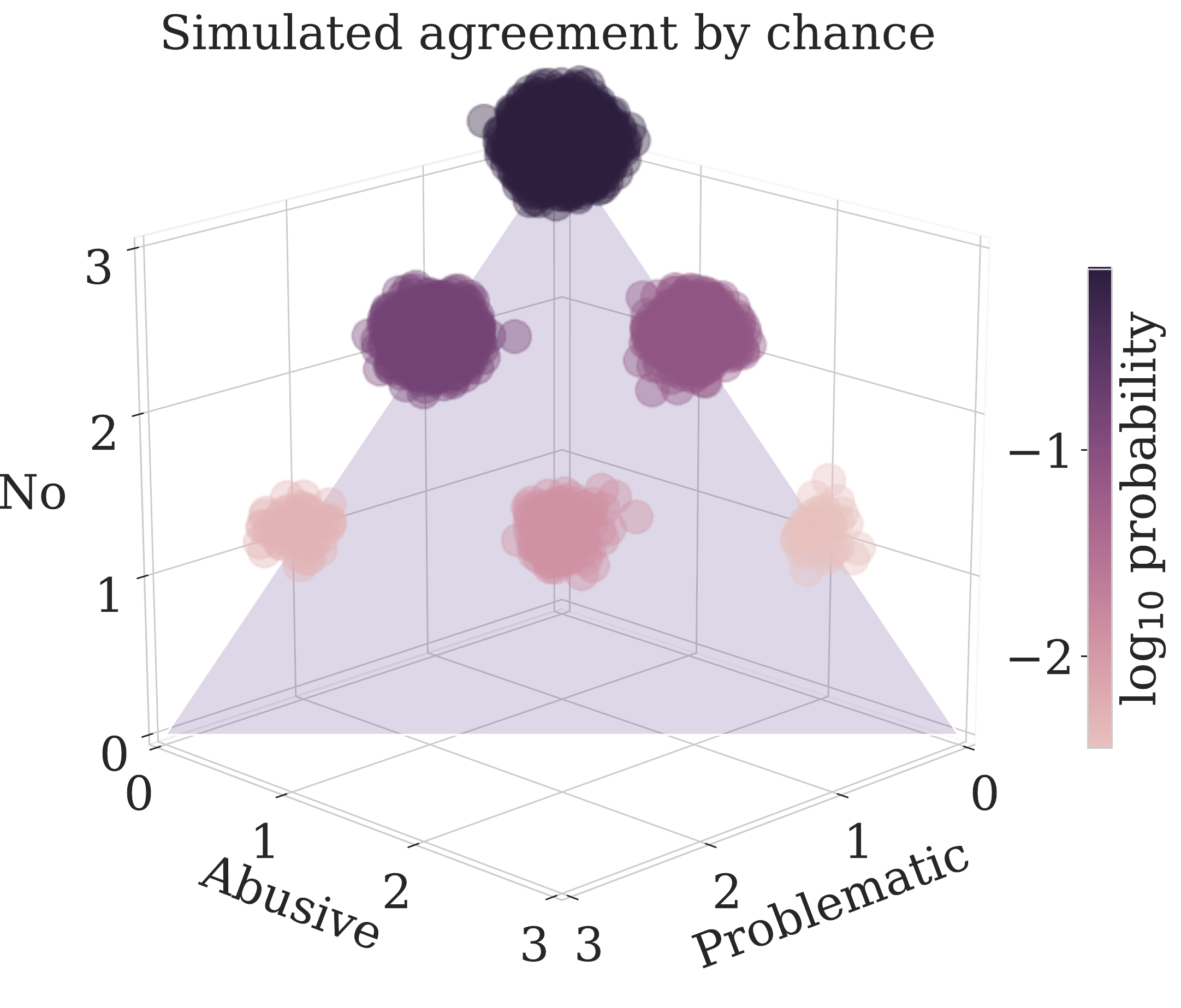}
        \label{fig:simplex:chance}
	}
	\caption{\small \label{fig:simplex}
	    Visualizing the distribution of annotations $a^{(t)}$ ($+$ jitter for clarity) in the multinomial 2-simplex.
	    The corners are events of \textit{complete} agreement. The center is \textit{no agreement} with the non-ordinal assumption, but partial agreement with ordinality.
	    Left to right:
	    Simulated perfect agreement, $a_c^{(t)}=3, c \sim \hat{P}(C)$;
	    Agreement among 3 experts on 1000 tweets: empirical probabilities are visually amplified by over-sampling $a^{(t)}$ to 20k;
	    Agreement among $N=3$ Decoders per tweet: if $N>3$, raters are chosen randomly;
	    Simulated agreement-by-chance only: $a^{(t)} \sim \mathit{Multinomial}(N=3, p=\hat{P}(C))$.
	    The multinomial assumes independence between trials.
	    A hierarchical modeling approach can capture inter-rater dependence \cite{Kalaitzis:discrete13, Silva:projections15}.
	}
\end{figure}

\textbf{Fleiss' kappa}:
A rater can annotate a tweet as class
$c \in C = \{\mathrm{No}, \mathrm{Problematic}, \mathrm{Abusive}\}$.
The annotation
$a = (a_{No}, a_{Pr}, a_{Ab}), a_c \in \{0,1,2,3\}, \Sigma_{c} a_c = N$,
contains the class counts for a tweet annotated by $N$ raters.
The overall agreement for a set of tweets $T$ is
$\kappa = \frac{1}{|T|} \Sigma_{t} \kappa^{(t)}$,
where
$\kappa^{(t)} = \frac{\Sigma_{c} r_c^{(t)} - \Sigma_{c} p_{c}^{2}}{1 - \Sigma_{c} p_{c}^{2}}$
is the \textit{within-tweet} agreement,
and
$r_c = \frac{a_c (a_c - 1)}{N (N - 1)} \in [0,1]$
is the fraction of pairs of raters that agree on $c$.
$\hat{P}(C=c) = p_c$
is the empirical probability of $c$,
hence
$\Sigma_{c} p_{c}^{2}$ is the probability of agreement-by-chance. \\

\begin{figure}
    \centering
    \adjincludegraphics[height=3.2cm, trim={{0\width} 0 0 0}, clip]
        {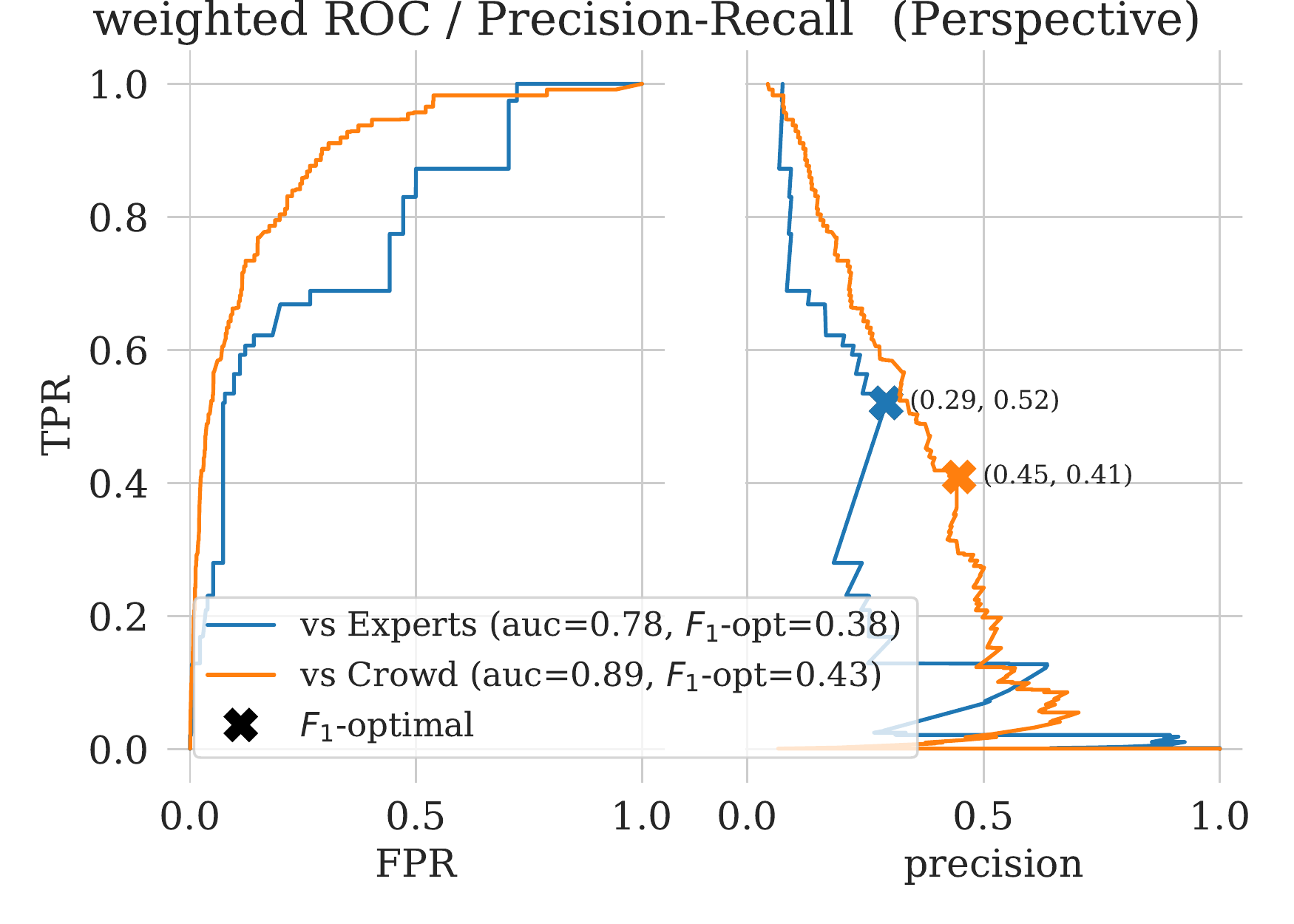}
    \vspace{-.2cm}
    \adjincludegraphics[height=3.2cm, trim={{0\width} 0 0 0}, clip]
        {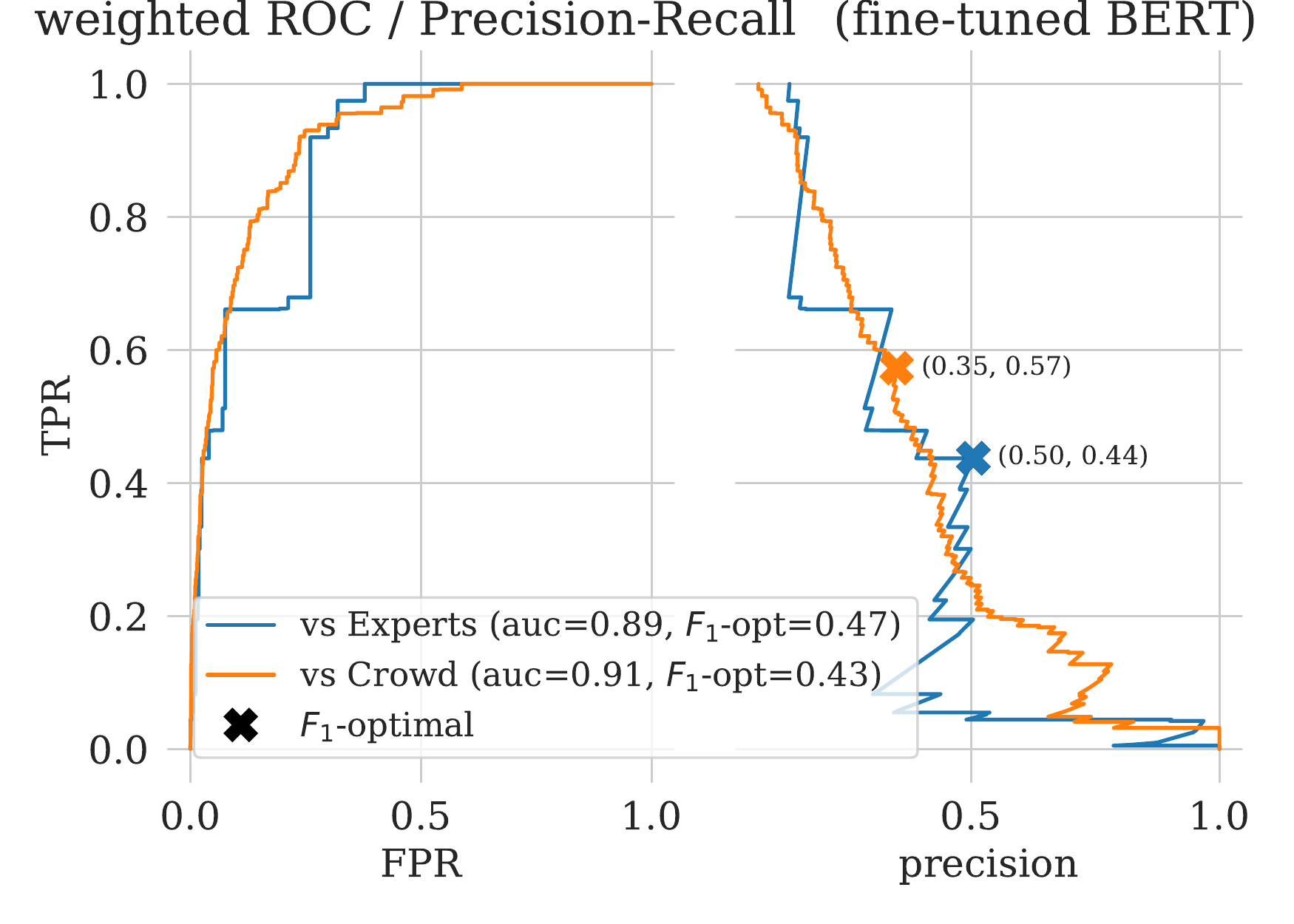}
    \adjincludegraphics[height=3.2cm, trim={{0\width} 0 0 0}, clip]
        {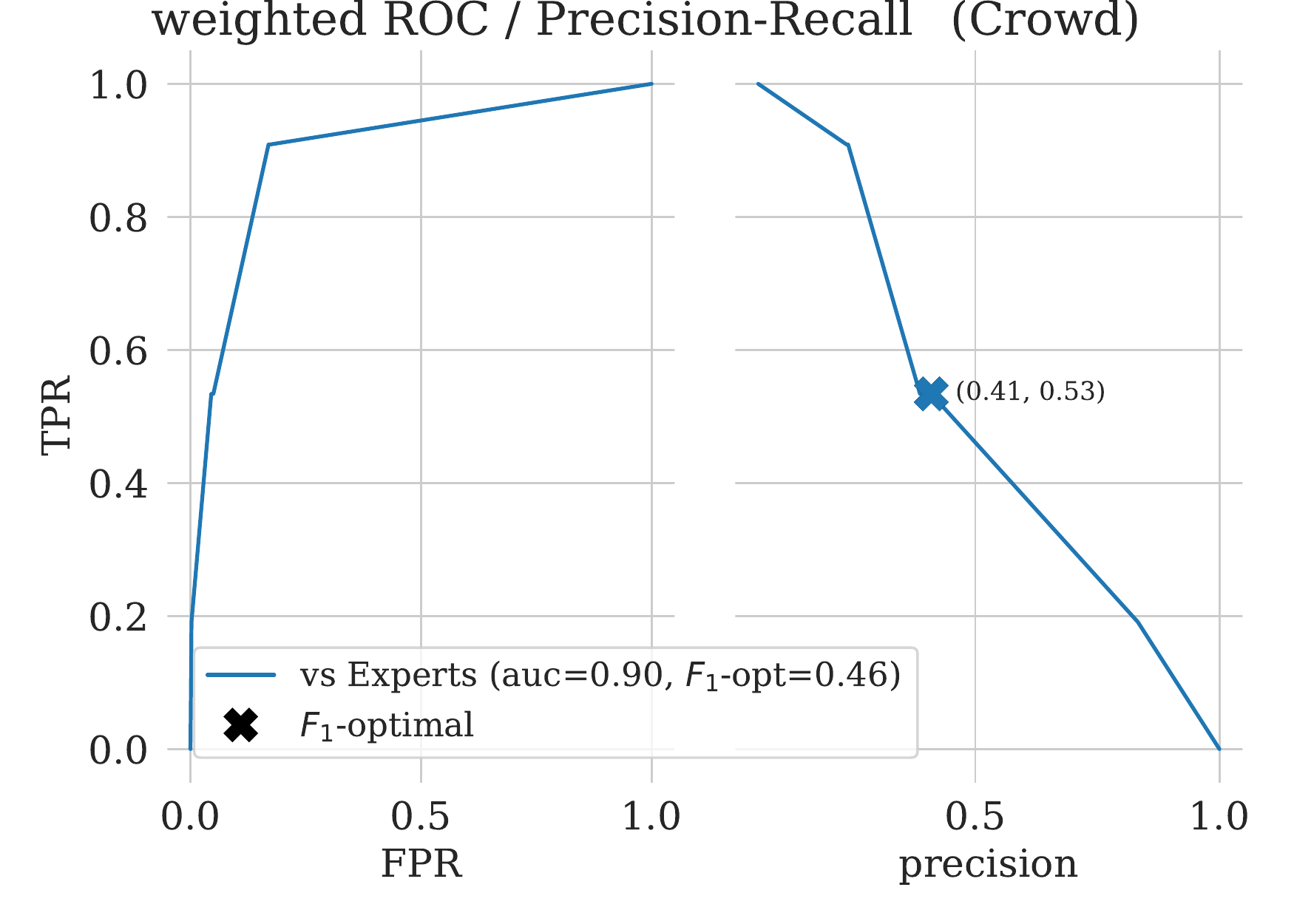}
    \caption{
        \small
        \textbf{Left \& center}: performance of Perspective API and fine-tuned BERT classifiers, with respect to the experts' and crowd's labels. \textbf{Right}: performance of the crowd-as-a-classifier against the experts' labels.
        \textbf{Note}: \textit{Recall} is equivalent to \textit{TPR}, hence the y-axes (TPR<>Recall) of the two plots are aligned.
        \label{fig:perf_perspecive_unweighted}
    }
    \label{fig:perf_perspecive_weighted}
\end{figure}

\textbf{ICC (intra-class correlation)}:
Let
$\mathbf{A} \in \mathbb{R}^{|T| \times N}$
be the matrix of annotations,
$A_{i,j} \in \{0,1,2\}$
(ordinal values raters can assign), and each row is a tweet annotated by $N$ random raters.
The tweet-specific mean is
$\mu_i=\frac{1}{N}\Sigma_j A_{i,j}$
and the overall mean is
$\mu=\frac{1}{|T|}\Sigma_{i} \mu_i$.
The \text{within}-tweet disagreement for tweet $i$ is
$V^{(i)} = \frac{1}{(N-1)} \Sigma_{j} ~ (A_{i,j} - \mu_i)^2$,
and its average  $V_w = \frac{1}{|T|} \Sigma_{i} V^{(i)}$ is the overall within-tweet variance.
Similarly, the \textit{between}-tweet variance is
$V_b = \frac{N}{|T|-1} \Sigma_i (\mu_i - \mu)^2$.
The $\mathrm{icc}$ can now be expressed in terms of a one-way ANOVA \cite{Shrout:icc79}:
$\mathrm{icc} = \frac{V_{b} - V_{w}}{V_b + (N-1)V_w}$,
the fraction of variation in annotations that is not explained by between-tweet disagreements.

\textbf{Results}: Table~\ref{tab:agreement} and Figure \ref{fig:simplex} (mid left / mid right) show more agreement among the experts than among the volunteers -- higher $\kappa$ and ICC among the former. There is also more agreement when assessing the presence of abuse than when assessing the type of abuse.

\begin{table}
\vspace{-3em}
\centering
\begin{minipage}[t]{0.34\textwidth}
\caption{\small Agreement per variable and per labeling cohort.}
    \label{tab:agreement}
    \centering
    \resizebox{1\textwidth}{!}{
        \begin{tabular}{lllll}
            \toprule
            Labels from & \multicolumn{2}{l}{Crowd} & \multicolumn{2}{l}{Experts} \\
            \midrule
            {}            & $\kappa$ &  ICC  & $\kappa$ &  ICC \\
            \midrule 
            Contain Abuse &    .26 &  .35   &     .54 &  .70   \\
            Type of Abuse &    .16 &    -   &     .74 &    -   \\
            \bottomrule
        \end{tabular}
    }
\end{minipage}\hfill%
\begin{minipage}[t]{0.60\textwidth}
 \caption{\small Classifier performance vs. crowd vs. expert labels}
            \label{tab:classifier-performance}
            \centering
            \resizebox{1\textwidth}{!}{
                \begin{tabular}{lllllllll}
                    \toprule
                    Labels from & \multicolumn{4}{l}{Crowd} & \multicolumn{4}{l}{Experts} \\
                    \midrule
                                & $Precision$ & $Recall$ & $F_1^*$ & $AP$ &  $Precision$ & $Recall$ & $F_1^*$ & $AP$ \\
                    \midrule
                    Naive Bayes     &  .13 &  .25 &  .17 &  .11 &  .40  &  .27 &  .32 &  .21 \\
                    Crimson Hexagon &  .14 &  .40 &  .20 &  -   &  .05 &  .04 &  .04 &  -   \\
                    Davidson et al. &  .53 &  .27 &  .36 &  .25 &  .35 &  .46 &  .39 &  .25 \\
                    Perspective API &  .45 &  .41 &  \textbf{.43} & .34 &  .29 &  .52 &  .38 &  .25 \\
                    Fine-tuned BERT &  .35 &  .57 &  \textbf{.43} & \textbf{.40} &  .50 &  .44 &  \textbf{.47} &  .36 \\
                    \midrule
                    Crowd           &  - &  - &  - &  - &  .41 &  .53 &  .46 &  \textbf{.39} \\
                    \bottomrule
                \end{tabular}
            }
\end{minipage}
\end{table}

\subsection{Comparison of baseline classifiers}

\label{sec:baseline}
The core focus in this study is to build and analyze the dataset, with a view to extend that analysis to the remaining $2$M unlabeled tweets using state of the art models. We prepare this follow-up research community effort by establishing baselines on various classification models.

\textbf{Classifiers}: In Table \ref{tab:classifier-performance}, Naive Bayes refers to the classifier from~\cite{Katia2017}. Crimson Hexagon refers to sentiment labels -- Category and Emotion -- from Crimson Hexagon. We also benchmarked the pre-trained classifier from Davidson et al. \cite{Davidson:hatespeech17}. Perspective API refers to the public toxicity scoring API provided by Jigsaw~\cite{PerspectiveAPI, Hosseini:deceiving17}. We also trained our own model, which combined a pre-trained BERT embedder \cite{devlin2018bert} and an abuse-specific embedding trained from scratch. For details see \ref{sec:model}.
    
\textbf{Methodology}:
For this analysis, we conflate the labels \textit{Problematic} and \textit{Abusive} into one positive (Abusive) class.
The crowd labels are the majority votes over labels on tweets labeled by exactly three volunteers.
The expert labels are majority votes over labels from the three domain experts mentioned in Section \ref{sec:data_collection}.
For Crimson Hexagon, we define Abusive as the intersection of \texttt{Category = Basic\ Negative} and \texttt{Emotion = Anger\ |\ Disgust}.

\textbf{Results}:
Table \ref{tab:classifier-performance} shows the $F_1^*$ (optimal $F_1$ score), corresponding precision and recall, and the Average Precision ($AP$), to evaluate several abuse detection classifiers with respect to labels from the crowd and from the experts.

\section{Discussion}  \label{sec:discussion}

\textbf{Dataset availability and reproducibility}: Amnesty International intends to publish as much of the dataset as possible to encourage replication and further research on the topic. At the very least the URLs of the tweets and the grades will be made public. Publishing the actual tweets is more delicate due to Twitter Terms and Conditions. Publishing the meta-data on the graders (gender, location) would be of great interest, but is still under discussion from an ethical point of view.

\textbf{Future work}: We aim to eventually apply different classifiers to the whole unlabeled dataset, so as to scale up the human rights researchers' work by sifting through the huge amount of tweets.
As shown in Section~\ref{sec:baseline}, this will require a careful tuning of models to increase the precision beyond its current performance.
In parallel, we also want to analyze the scale, typology and intersectionality of abuse, either on the labeled set or on the classified extra $2$M tweets, for the planned media campaign.

 \textbf{Social impact}: The sheer volume of hateful speech on social media has recently prompted governments to put strong pressure on social media companies to remove such speech \citep{Gambck2017UsingCN}. The moderation of abusive messages at scale requires some form of automated assistance. Our results highlight the double challenge of automatic abuse classification: the subjectivity in the labels and the limited ability of state-of-the-art classifiers to generalize beyond training data. This all points toward the need for systems where human subtlety and context awareness are empowered by automatic pre-screening.

Whether the companies themselves should be trusted with (or required to implement) such moderation, or whether they should fund or be supervised by a third-party neutral watchdog, goes far beyond a purely technical conversation. This is why collaboration between technical experts (machine learners, data scientists) and domain experts (human rights researchers, anti-censorship activist, etc.), as well as society in a broader sense, is so important for genuinely impactful AI for Social Good efforts.

\section*{Acknowledgements}
\vspace{-1em}
{\small
    We are extremely grateful to all the volunteers from Amnesty Decoders for their hard work.
    We would like to thank Nasrin Baratalipour, Francis Duplessis, Rusheel Shahani and Andrei Ungur for modelling support.
    We also thank Jerome Pasquero for his support, and the Perspective team for access to their API.
}

\bibliographystyle{unsrtnat}
\bibliography{references.bib}

\newpage
\appendix

\section{Population definition}\label{sec:population}
We selected politicians and journalists with an active, non protected Twitter account, with fewer than 1 million followers.     The group included: 
\begin{itemize}
    \item  All women members of the British Parliament (220, including 22 women who left parliament during the June 2017 elections and excluding one politician with over 1 million followers); 
\item All women in the United States Congress (107, excluding 3 politicians with more than 1 million followers); 
\item And women journalists working at the following news organizations,  selected to represent a diversity of media bias:
\begin{itemize}
    \item Breitbart (16),
    \item Daily Mail (78),
    \item The Sun (54),
    \item The Guardian (124),
    \item The New York Times (278),
    \item Gal-Dem (23),
    \item PinkNews (9).
    \end{itemize}
\end{itemize}

\section{Agreement analysis} \label{appendix:agreement}
    \subsection{Fleiss' Kappa} \label{appendix:kappa}
        \paragraph{Notation}
        A rater can annotate a tweet as class
        $~c \in C = \{\mathrm{No}, ~\mathrm{Problematic}, ~\mathrm{Abusive}\}$.
        The annotation tuple
        $a = (a_{No}, ~a_{Pr}, ~a_{Ab}), ~~ a_c \in \{0,1,2,\dots\}, ~~ \Sigma_{c} a_c = N$,
        contains the class-specific counts for a tweet annotated by $N$ raters.
        
        \paragraph{Estimation of $\kappa$}
        The \textit{within-class} agreement-ratio
        $r_c = \frac{a_c ~ (a_c - 1)}{N ~ (N - 1)} \in [0,1]$
        is the ratio of pairs of raters that agree on $c$, over the total of pairs of $N$ raters.
        $\hat{P}(C=c) = p_c$ is the empirical marginal probability of class $c$.
        Hence, $\Sigma^{}_{c} ~ p_{c}^{2}$ is the overall probability of agreement by chance across a dataset of tweets.
        For a specific tweet $t$ we can compute the \textit{within-class} agreement
        $\kappa_{c}^{(t)} = \frac{r_c^{(t)} - p_{c}^{2}}{1 ~-~ \Sigma_{c} p_{c}^{2}}$,
        where the numerator is the \textit{agreement-above-chance} attained on $c$,
        and the denominator is the \textit{best-case-scenario} (maximal) agreement-above-chance attainable across classes.
        Hence $\kappa_c^{(t)}$ is the fraction that the attained agreement in $c$ contributes to the best-case scenario, while accounting for agreement-by-chance.
        Finally, the \textit{within-tweet} agreement for a tweet $t$ is the sum across classes,
        $\kappa^{(t)} = \Sigma_{c} \kappa_{c}^{(t)}$,
        and the overall agreement across a set of tweets $T$ is the expectation
        \begin{equation}
            \kappa = \mathbb{E}_{T}[\kappa^{(.)}] \approx \tfrac{1}{|T|} \Sigma_{t} \kappa^{(t)}.
        \end{equation}

    \subsection{ICC (intra-class correlation)}
    
        \paragraph{Notation}
        We denote the matrix of annotation as $\mathbf{A} \in \mathbb{R}^{|T| ~\times~ N}$.
        In this work, $A_{i,~j} \in \{0,1,2\}$ (ordinal values raters can assign), where each row represents a tweet annotated by $N$ random raters.
        \paragraph{Algorithm}
        The tweet-specific mean is
        $\mu_i=\frac{1}{N}\Sigma_j A_{i,j}$
        and the overall mean is
        $\mu=\frac{1}{|T|}\Sigma_{i} \mu_i$.
        We can express the \textit{within-tweet} disagreement as the \text{within-tweet variance}
        $V^{(i)} = \frac{1}{(N-1)} \Sigma_{j} ~ (A_{i,j} - \mu_i)^2$.
        Then the average of within-tweet disagreements expresses the overall within-tweet variance, $V_w = \frac{1}{|T|} \Sigma_{i} V^{(i)}$.
        Similarly, the \textit{between-tweet variance} is $V_b = \frac{N}{|T|-1} \Sigma_i (\mu_i - \mu)^2$.
        Note that the $i$-th tweet is \textit{polarized} when $V^{(i)}$ is maximized, i.e. half of the raters choose 0 and the other half choose 2.
        In the extreme scenario that all tweets are maximally polarizing, $V_b = 0$.
        Therefore $V_b$ expresses the overall tendency for \textit{disagreement-by-chance}.
        All classes of ICC are equivalent to a type of ANOVA (\textit{ANalysis Of VAriance}) in linear mixed-effects model of annotations \cite{Shrout:icc79}.
        In our case,
        \begin{equation}
            \mathrm{icc}(1,k)=\frac{V_{b} - V_{w}}{V_b + (N-1)V_w}
        \end{equation}
        Intuitively, the ANOVA framework defines agreement as the fraction of variation in annotations that is not explained by between-tweet disagreements.

        \paragraph{Systemic disagreement}
        As mentioned above, in extreme scenarios where $V_b$ is small, the ICC can be negative. 
        Negative $\kappa$ and $\mathrm{icc}$ values might seem like an artifact of degenerate or extreme data, only to be dismissed as \textit{no agreement} in the downstream analysis. At closer inspection, the numerator shows that subtracting the agreement-by-chance yields a measure of systemic disagreement: e.g.\ expecting $P(\textrm{agreement-by-chance}) = 0.9$ but observing agreement only 20\% of the time, implies a systemic cause for polarizing opinions (e.g.\ controversial content, raters annotating with different rules).
    

\section{Importance sampling analysis}\label{sec:reweight}
    \paragraph{Population and Crimson sets $W$ and $C$:}
    We denote the population (\textit{World set}) as $W$, and the sample obtained from the Twitter firehose (\textit{Crimson set}) as $C$.
    Members of the sets $W$ and $C$ are observation tuples $(t,k,d)$,
    where $t$ is the text content of a tweet,
    $k \in \{0,1\}$ is the output of a Naive Bayes Classifier $\mathrm{NBC}: t \mapsto k$,
    and $d$ is the day in 2017 that a tweet was published:
    \begin{align}
        C \subset W = \{(t,k,d)\}
    \end{align}
    \paragraph{Distributions $p_W$ and $p_C$:}
    We define $p_W$ and $p_C$ as the probability mass over sets $W$ and $C$, respectively, and any marginals and conditionals thereof:
    \begin{align}
        p_{W}(t,k,d) &= p_{W}(t,k|d) ~p_{W}(d) \\
        p_{C}(t,k,d) &= p_{C}(t,k|d) ~p_{C}(d) \label{eq:proba_c}
    \end{align}
    The density $p_W(d)$ is directly available from the daily total volumes $n_d$ of tweets matching the query,
    total that is provided by Crimson Hexagon alongside the smaller sampled set  $C$:
    \begin{gather}
        n_d = |\{(t,k,d') \in W : d' = d\}| \text{ provided as metadata,} \notag \\
        p_W(d) = \frac{n_d}{\sum_{d'} n_{d'}}\,.
        \label{eq:p_w_d}
    \end{gather} 
    The \textit{Crimson set} $C$ is constructed by uniform sampling over tweets in $W$, such that for any day $d$, the conditional probabilities over both sets are equal:
    \begin{equation} \label{eq:proba_c_cond_d}
        p_{C}(t,k|d) = p_{W}(t,k|d)
    \end{equation}
    Then, using eq. (\ref{eq:proba_c_cond_d}) in (\ref{eq:proba_c}):
    \begin{equation}
       p_{C}(t,k,d) = p_{W}(t,k|d) ~p_{C}(d)
    \end{equation}
    
    \paragraph{Constructed set $A$:}
    The final set  $A$ is defined as the union
    \begin{align}
        A = B \cup F, \label{eq:amnesty_set}
    \end{align}
    where
    \begin{align}
        B &= \{(t,k,d) \sim p_{B}(t,k,d) \simeq p_{W}(t,k|d) ~\hat{p}_{W}(d) \simeq ~p_{W}(t,k,d) \}\
    \end{align}
    approximates the world joint distribution through stratified sampling per day, and 
    \begin{align}
        F &= \{(t,k,d) \in C \backslash B: k=1 \}
    \end{align}
    is an enriched sample resulting from pre-filtering by a simple Naive Bayes classifier. The cardinalities of these sets are: $|C| = 2.2M$, $|B| = 215k$, $|F| = 60k$ and $|A| = 275k$. \\
    With $~\beta = \frac{|F|}{|B| + |F|}$, and $z(d)$ a normalizing constant depending on $d$: 
    \begin{equation} \label{eq:pA_cond_d}
        p_A(t,k|d) = \frac{\beta \mathbb{I}(k=1) ~p_A(t,k|d) + (1-\beta) ~p_A(t,k|d)}{z(d)}
    \end{equation}
    where $\mathbb{I}(.)$ is the indicator function. \\
    The conditional probabilities of a tweet are identical in $W$ and $A$:
    \begin{equation} \label{eq:t_cond}
        p_W(t|k,d) = p_A(t|k,d) \,.
    \end{equation}
    
    Combining equations (\ref{eq:t_cond}) and (\ref{eq:pA_cond_d}) leads to:
    \begin{equation}
        p_A(t,k|d) \propto
        \beta \mathbb{I}(k=1) ~p_W(t|k,d) ~p_W(k|d) + (1-\beta) ~p_W(t|k,d) ~p_W(k|d) \,.
    \end{equation}
    
    \paragraph{Importance weights $w_i$:}
    Estimating statistics on the world set $W$ using the samples in set $A$ can be achieved using importance sampling, i.e. assigning a specific weight $p_W(t,k,d)/p_A(t,k,d)$ to each triplet $(t,k,d)$ in $A$. 
    
    For each tweet $(t,k,d) \in A$, we define the weighting function $w$
    \begin{align} 
            w(t,k,d) &= \frac{p_W(t,k, d)}{p_A(t,k, d)}\notag \\
                &= \frac{p_W(t|k,d) ~p_W(k, d)}
                        {p_A(t|k,d) ~p_A(k, d)}\,.
                        \label{eq:weights_def}
    \end{align}
    Injecting equation (\ref{eq:t_cond}) in equation (\ref{eq:weights_def}), we can simplify by $p_W(t|k,d)$:
    \begin{align}
        w(t,k,d) &= \frac{p_W(k,d)}
                    {p_A(k,d)} \notag \\
          &= \frac{p_W(k|d) ~p_W(d)}
                    {p_A(k|d) ~p_A(d)} \label{eq:weights_to_evaluate}\,.
    \end{align}
    
    Since $A$ is a finite set, the probability mass functions $p_A(k|d)$ and $p_A(d)$ in equation~(\ref{eq:weights_to_evaluate}) are directly accessible by simple counting. 
    
    The probability mass functions $p_W(d)$ is known from~(\ref{eq:p_w_d}). The term $p_W(k|d)$ is not available in closed form, but can be estimated straightforwardly. Indeed from equation~(\ref{eq:proba_c_cond_d}) we have 
    $p_W(k|d) = p_C(k|d)$, and the latter can be estimated by simple counting on $C$, leading to empirical estimate:
    \begin{align}
    \hat p_W(k|d)  = \frac{|\{(t,k',d') \in C : k'=k, d'=d\}|}{|\{(t,k',d') \in C : k'=k\}|} \label{eq:hat_p_w_k_d}\,.
    \end{align}
    This leads to the final plug-in estimator of the importance weights:
    \begin{align}
    \hat w(t,k,d) = \frac{\hat p_W(k|d) p_W(d)}{p_A(k|d) p_A(d)}\,.
    \end{align}
    
    For any given function $f(t,k,d)$, we therefore estimate its expectation in the whole population $W$ using the self-normalized importance estimator: 
    \begin{equation}
        \hat{\mathbb{E}}_W [f(t,k,d)] = \sum_{(t_i, k_i, d_i) \in A}
                                        \frac{w_i}{\sum_j w_j}
                                        ~f(t_i, k_i, d_i)\,.
    \end{equation}
    where for any tweet $(t_i,k_i,d_i)$ with GUID (Globally Unique Identifier) $i$ we use the estimated unnormalized importance weight $w_i := \hat w(t_i,k_i,d_i)$.
    
    Note that for full mathematical rigour, the asymptotic consistency of the importance sampling estimator $\hat {\mathbb{E}}_W$ could be proven by showing that the replacement of the density estimator $\hat p_W$  in the plug-in estimator $\hat w$ is asymptotically valid. Such a proof could proceed along the lines of \cite{douc:moulines:2008}, but is outside of the scope of this article.
    \newpage

    
    
    
    


\section{Labeling tool screenshots}\label{sec:screenshots}
\begin{figure}[htbp]
    \begin{center}\includegraphics[width=.8\textwidth]{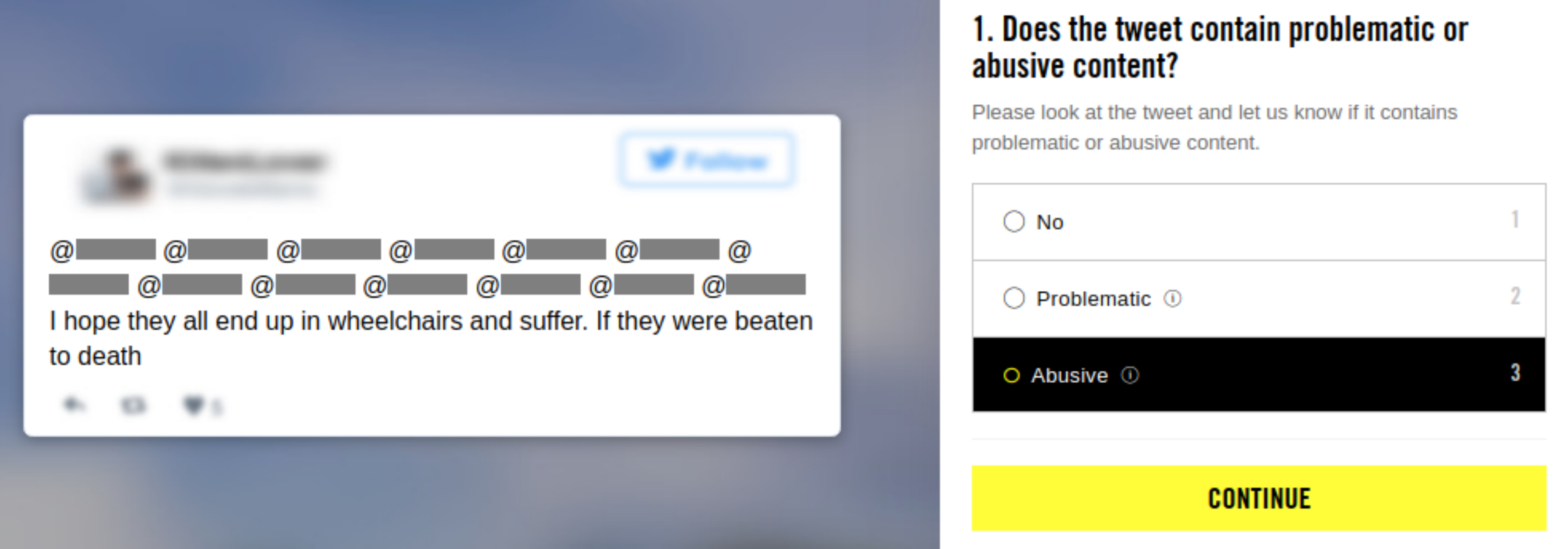}\end{center}
    \caption{\small First stage of labeling: Initial screen showing an anonymized tweet, with anonymized handles and first question.\label{fig:screenshot}}
\end{figure}

\begin{figure}[htbp]
    \centering
    \subfigure[\small Second stage of labeling: Identification of the type of abuse.]{%
    \includegraphics[width=.30\textwidth]{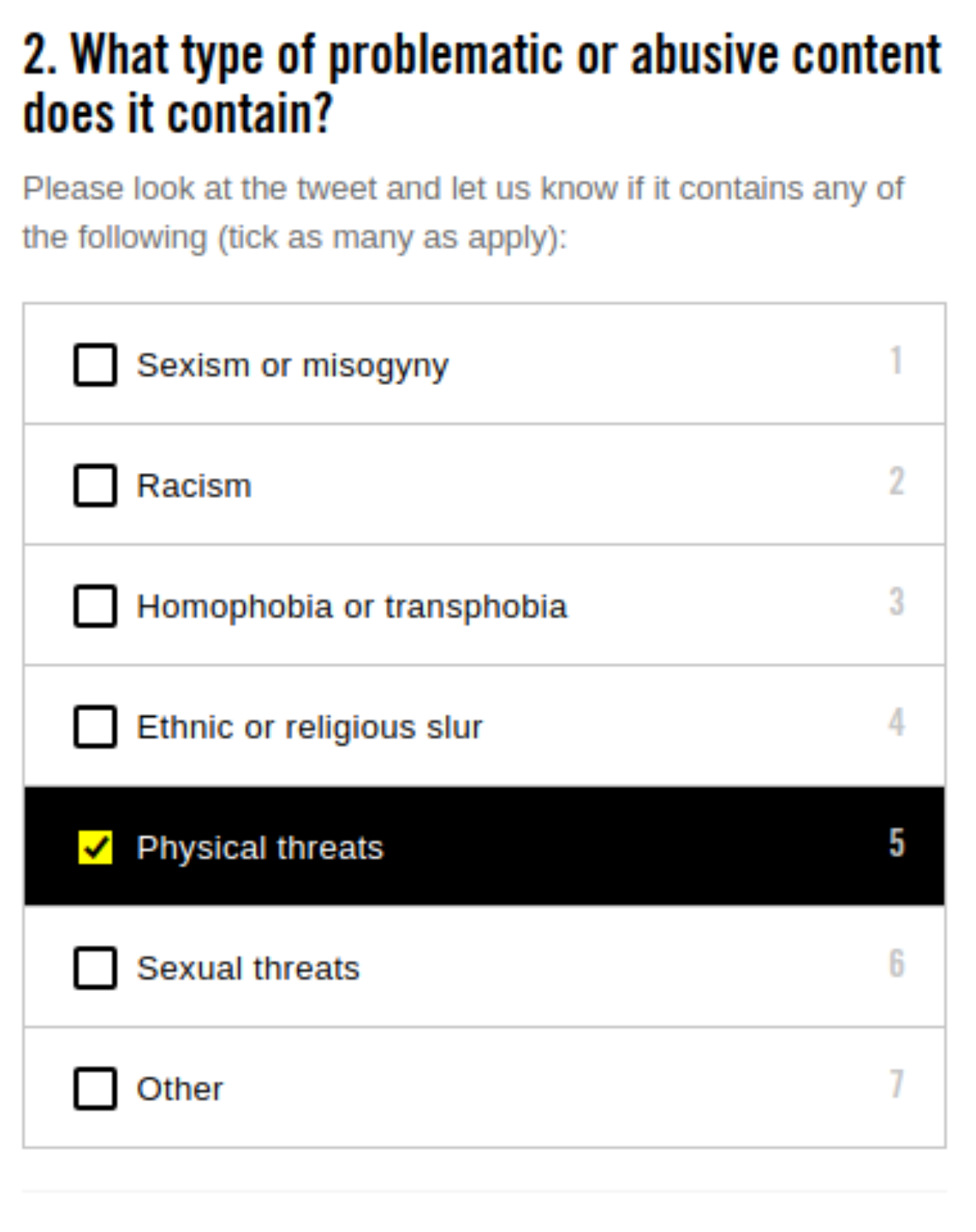}%
    }\hfill
    \subfigure[\small Third, optional stage of labeling: Identification of the part of the tweet that carries the abuse.]{%
    \includegraphics[width=.30\textwidth]{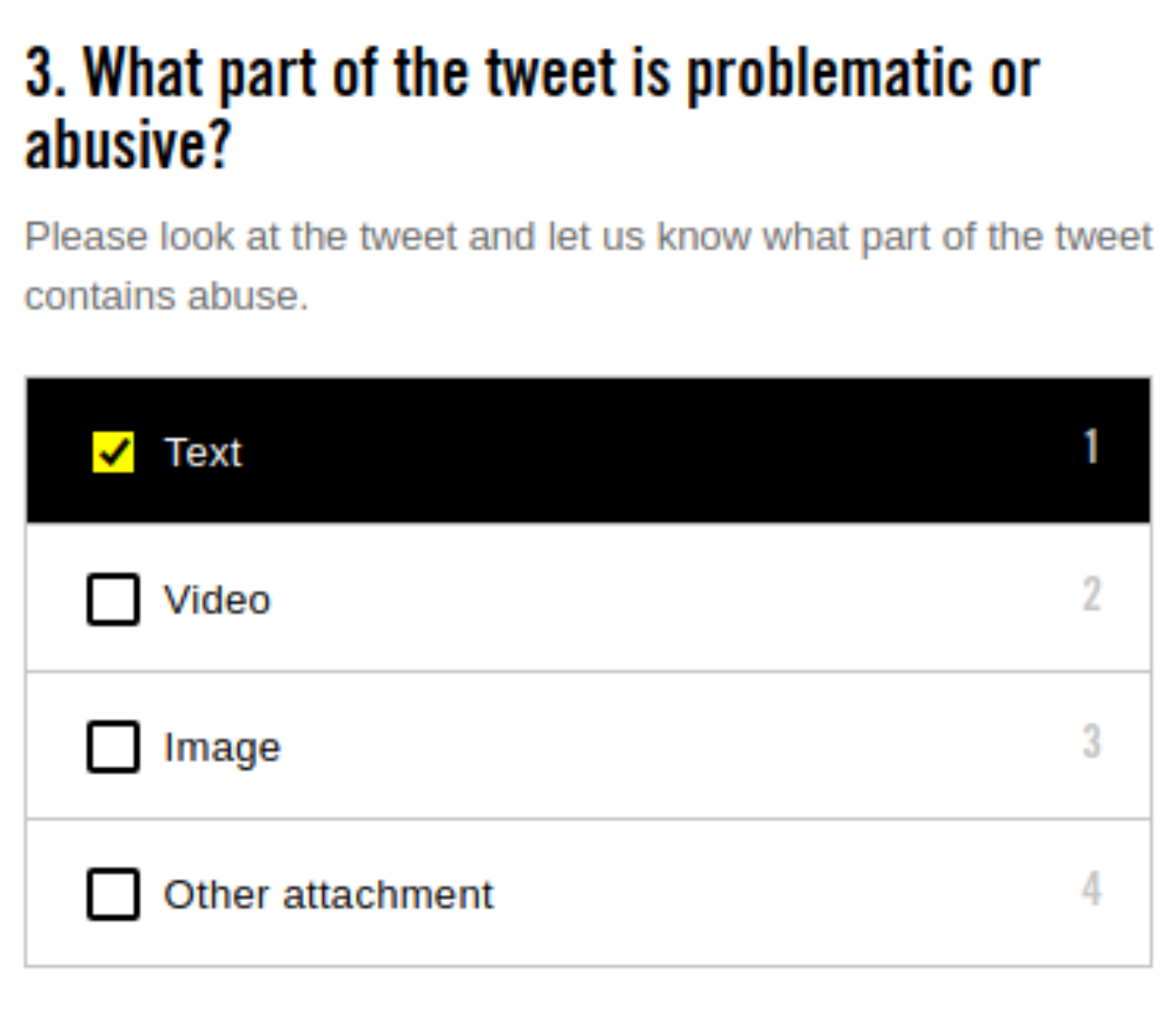}%
    }\hfill
    \subfigure[\small Warning displayed after classifying a tweet as abusive, to minimize the impact on the labelers' mental health.]{%
    \includegraphics[width=.30\textwidth]{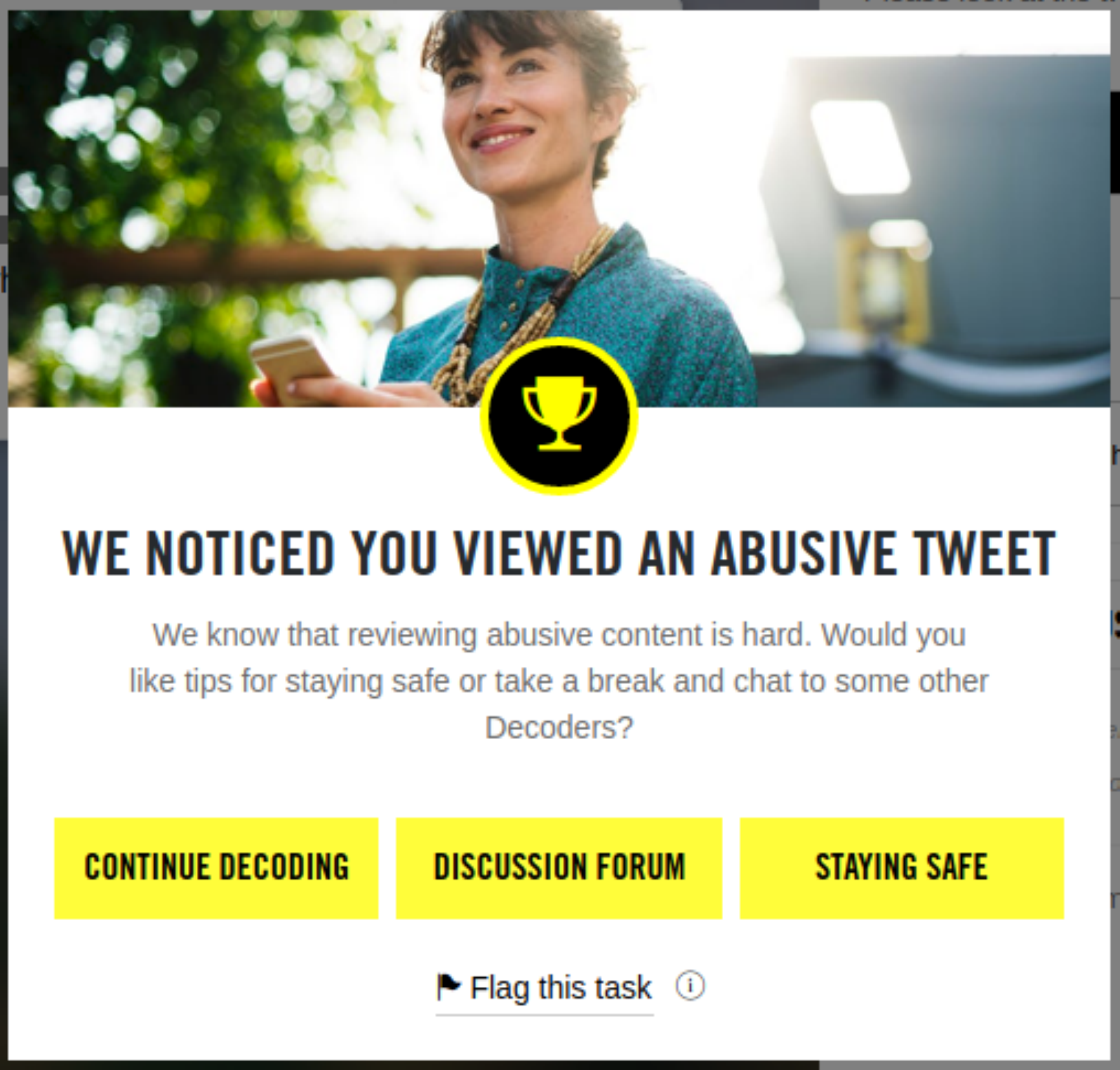}%
    }%
    \caption{\small Follow-up stages, conditional on the first stage of labeling.\label{fig:screenshot2}}
\end{figure}

The workflow presented to each grader by the labelling tool is illustrated in Figure~\ref{fig:screenshot} and Figure~\ref{fig:screenshot2}.
    
\section{Definitions and examples used in Troll Patrol - Trigger Warning} \label{sec:def_abuse}

    \paragraph{Abusive content}
    Abusive content violates Twitter’s own rules and includes tweets that promote violence against or threaten people based on their race, ethnicity, national origin, sexual orientation, gender, gender identity, religious affiliation, age, disability, or serious disease. 
    
    Examples include physical or sexual threats, wishes for the physical harm or death, reference to violent events, behaviour that incites fear or repeated slurs, epithets, racist and sexist tropes, or other content that degrades someone. For more information, see Twitter’s hateful conduct policy.
    
    In examples shown below, tweets were anonymized and only show  a standard template (incl. the author handle, the author profile picture, the tweet date and time, likes and retweets).

    \includegraphics[width=.5\textwidth]{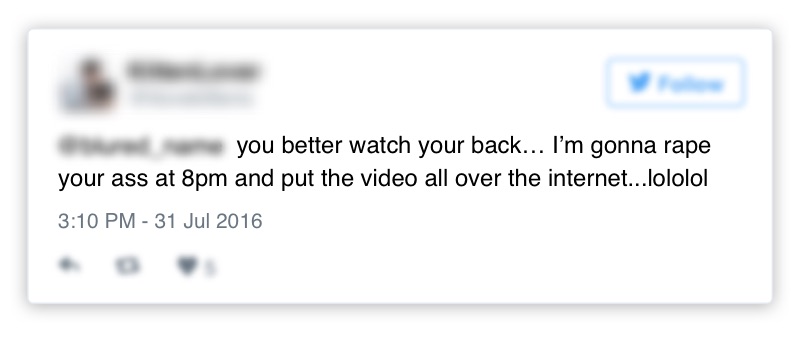}%
    \includegraphics[width=.5\textwidth]{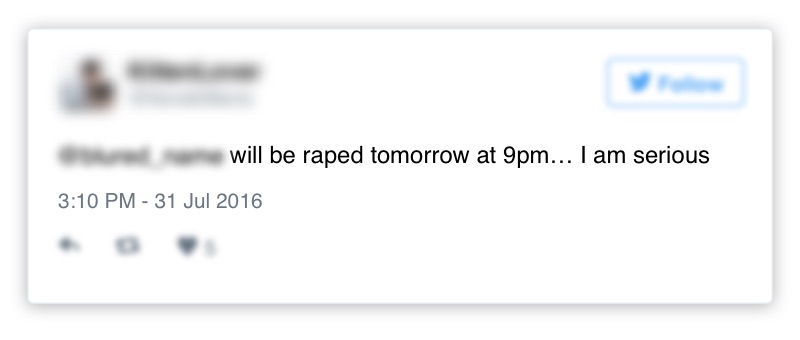}
    
    \paragraph{Problematic content}
    Hurtful or hostile content, especially if it were repeated to an individual on multiple or cumulative occasions, but not as intense as an abusive tweet.  It can reinforce negative or harmful stereotypes against a group of individuals (e.g. negative stereotypes about a race or people who follow a certain religion). Such tweets may have the effect of silencing an individual or groups of individuals.
    
    \begin{center}\includegraphics[width=.5\textwidth]{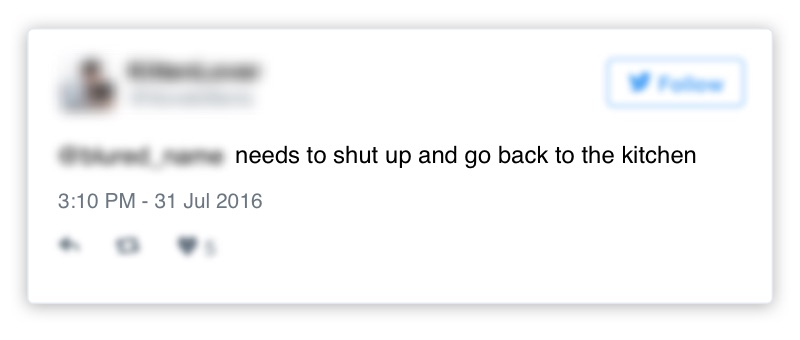}\end{center}
    
    \paragraph{Sexism or misogyny}
    Insulting or abusive content directed at women based on their gender, including content intended to shame, intimidate or degrade women. It can include profanity, threats, slurs and insulting epithets. 
    
    \includegraphics[width=.5\textwidth]{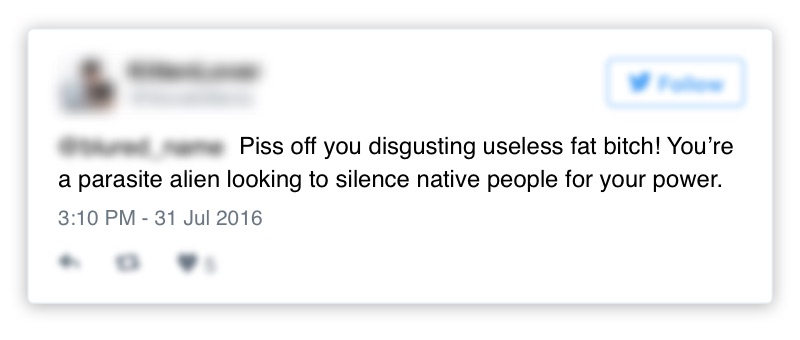}%
    \includegraphics[width=.5\textwidth]{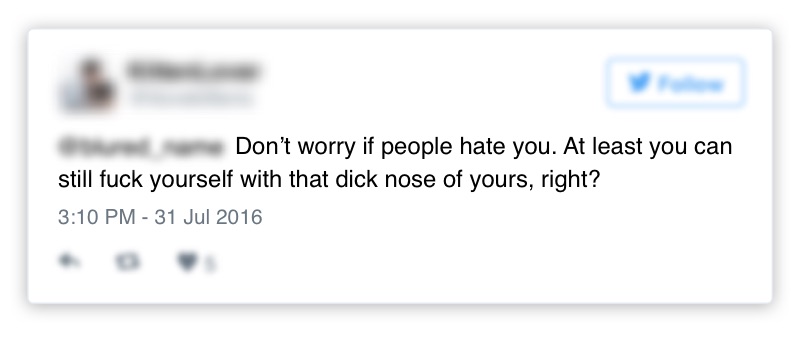}
    
    \paragraph{Racism}
    Discriminatory, offensive or insulting content directed at a woman based on her race, including content that aims to attack, harm, belittle, humiliate or undermine her.  
    
    \includegraphics[width=.5\textwidth]{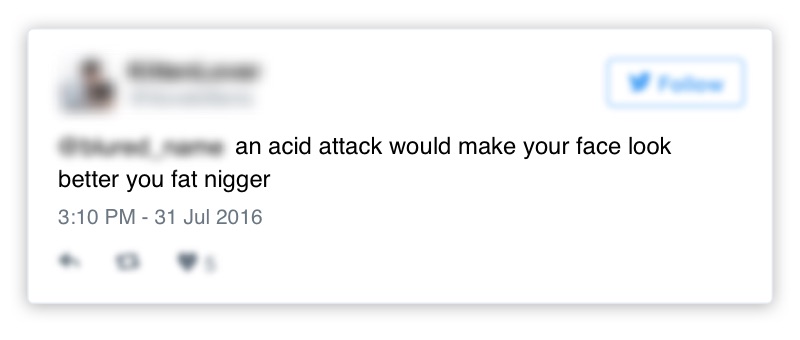}%
    \includegraphics[width=.5\textwidth]{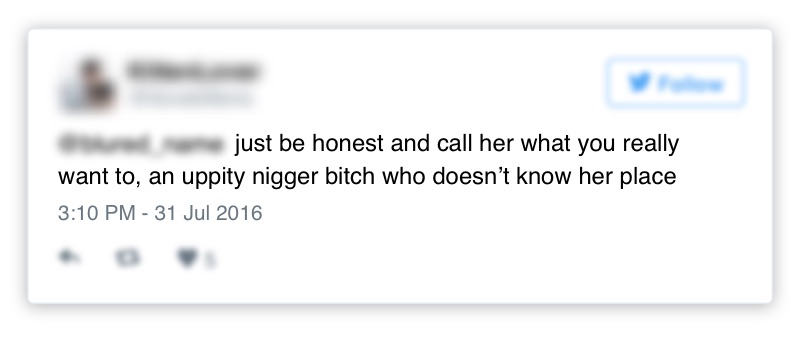}
    
    \paragraph{Homophobia or transphobia}
    Discriminatory, offensive or insulting content directed at a woman based on her sexual orientation, gender identity or gender expression. This includes negative comments towards bisexual, homosexual and transgender people. 
    
    \begin{center}\includegraphics[width=.5\textwidth]{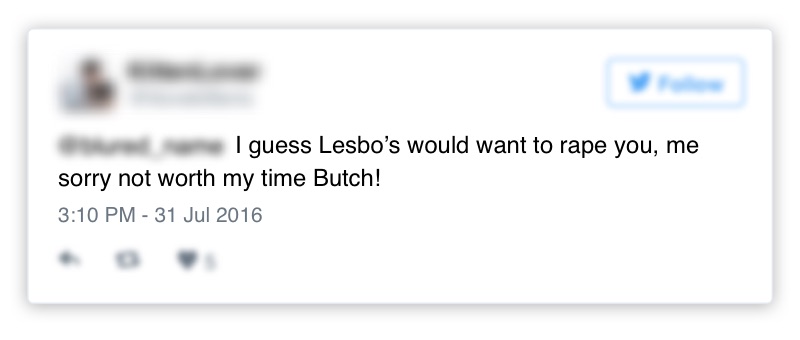}\end{center}
    
    \paragraph{Ethnic or religious slur}
    Discriminatory, offensive or insulting content directed at a woman based on her ethnic or religious identities.
    
    \includegraphics[width=.5\textwidth]{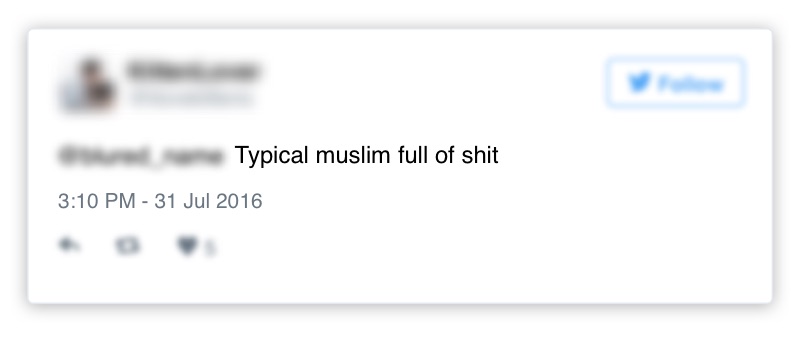}%
    \includegraphics[width=.5\textwidth]{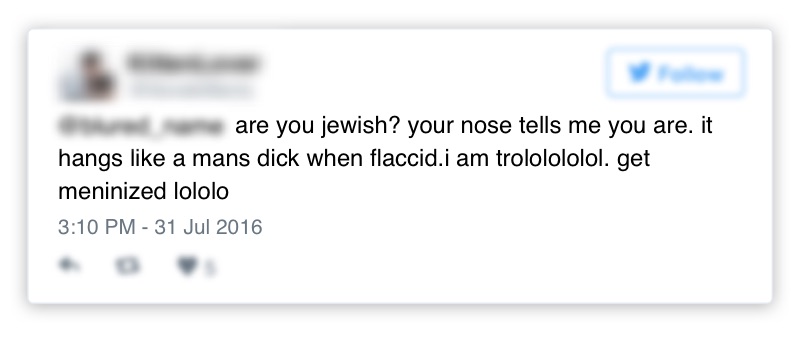}
    
    \paragraph{Physical threats}
    Direct or indirect threats of physical violence or wishes for serious physical harm, death, or disease. 
    
    \includegraphics[width=.5\textwidth]{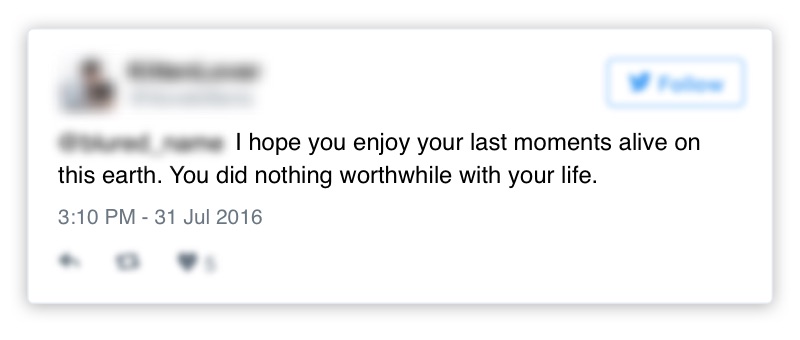}%
    \includegraphics[width=.5\textwidth]{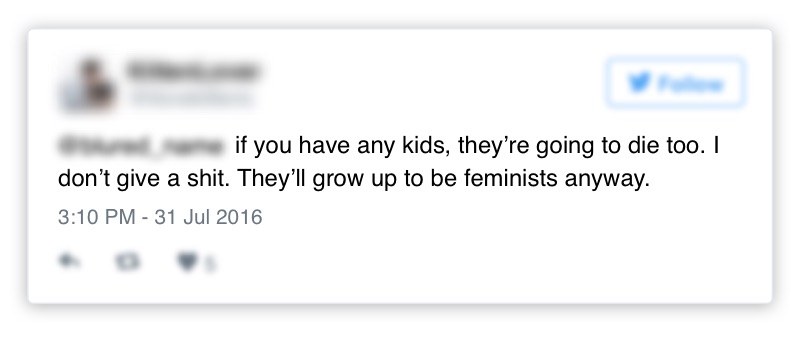}
    
    \paragraph{Sexual threats}
    Direct or indirect threats of sexual violence or wishes for rape or other forms of sexual assault.
    
    \includegraphics[width=.5\textwidth]{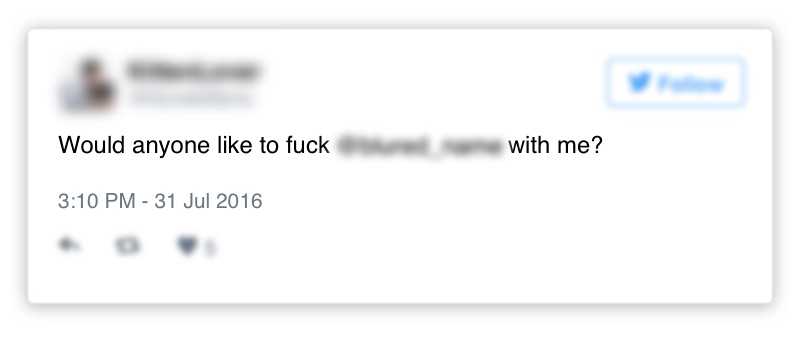}%
    \includegraphics[width=.5\textwidth]{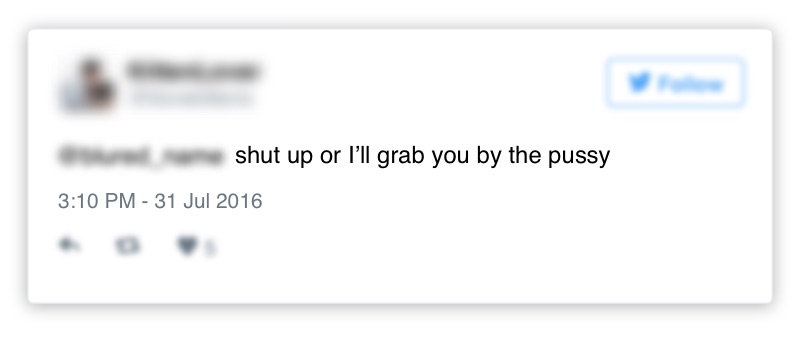}
    
    \paragraph{Other} 
    There will be some tweets that fall under the ‘other category’ that are problematic and/or abusive. For example, statements that target a user’s disability, be it physical or mental, or content that attacks a woman’s nationality, health status, legal status, employment, etc.
    
\section{Abuse Classification Model} \label{sec:model}
     \paragraph{Model architecture}
    We used a pretrained BERT model \citep{devlin2018bert} (12 layers, 768 units per layer) as the basis for our classification model. We took the final-layer representation of the first token in the sequence as a fixed-length tweet embedding (see Figure 3 of \citep{devlin2018bert}). The model was implemented in Pytorch, and made use of the BERT implementation provided at \citep{BERT}, which in turn utilizes a pre-trained model provided by Google.
   
    To account for out-of-vocabulary abusive words, we added a second single-layer word embedding (128 units), which we trained from scratch with a limited abusive vocabulary. This vocabulary included a list of ~1300 ‘possibly abusive’ words available online \citep{AbusiveWords}, and the 1000 words which occurred most disproportionately in the abusive class of the training data. To obtain a fixed-length representation from this embedder, we took the mean across words in each tweet.

    We concatenated these two representations to obtain a fixed-length tweet representation (of length 896), and passed througha fully-connected layer of 64 units before returning a decision via a binary softmax layer.
    
    \begin{figure}[hbt!]
     \centering
    \includegraphics[width=0.8\textwidth, clip]{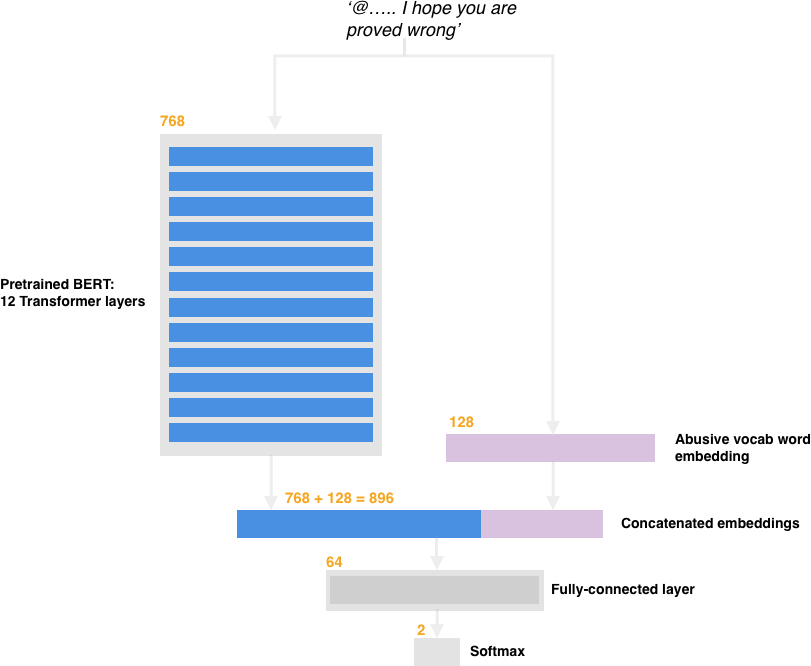}
    \caption{\small We combined a pre-trained BERT model with a word embedding exclusively including abusive words. The two embeddings were concatenated and passed through a fully-connected layer before a softmax layer returned the prediction. Numbers in orange are layer widths.}
    \end{figure}
    
    \paragraph{Data}
    We split the crowd-sourced data into train, validation, and test sets (90 : 5 : 5).
    We adjust all reported performance metrics for the original importance sampling (see \ref{sec:reweight}).
    
    \paragraph{Training}
    We trained the model end-to-end with stochastic gradient descent (learning rate = 0.0001, momentum=0.9) for 11 epochs, minimizing a cross-entropy loss.

\end{document}